\documentclass[11pt,a4paper]{article}
\usepackage[hyperref]{emnlp-ijcnlp-2019}
\usepackage{times}
\usepackage{latexsym}

\usepackage{algpseudocode}
\usepackage{algorithm}
\usepackage{amsmath}
\usepackage{amssymb}
\usepackage{hyperref}

\usepackage{url}

\aclfinalcopy 

\title{Fact-Checking at Scale with DimensionRank}

\author{Greg Coppola \\
Think different, again. \\
Founder \\
{\em greg@thinkdifferentagain.art } \\
\\
October 20, 2020
}

\date{}

\begin{document}
\maketitle
\begin{abstract}
The most important problem that has emerged after twenty years of popular internet usage is that of {\em fact-checking at scale}.
This problem is experienced acutely in both of the major internet application platform types, {\em web search} and {\em social media}.

We offer a working definition of what a ``platform'' is.
We critically deconstruct what we call the ``PolitiFact'' model of fact checking, and show it to be inherently inferior for fact-checking at scale to a {\em platform-based solution}. 

Our central contribution is to show how to effectively {\em platformize the problem of fact-checking at scale}.
We show how a two-dimensional rating system, with dimensions {\em agreement} and {\em hotness} allows us to create information-seeking queries not possible with the one-dimensional rating system predominating on existing platforms.
And, we show that, underlying our user-friendly user-interface, lies a system that allows the creation of formal proofs in the {\em propositional calculus}.

Our algorithm is implemented in our open-source {\em DimensionRank} software package available at \mbox{\em thinkdifferentagain.art}.
\end{abstract}

\section{Introduction}
{\em Web Search} and {\em Social Media} are the two most important internet applications.
Over the past 20 years, the primary problems of collecting, storing, and distributing messages and other information at scale have been solved.
A new problem, however, has emerged as a result of web search and social media.
This is the problem of {\em fact-checking at scale}.

Communications platforms have lowered the barriers to entry for independent news publications.
So there are now many competing {\em sources of truth}.
There has been a major loss in trust in the traditional, print and television-based publications.
And, publications have taken to being openly partisan, leading to a notion of {\em competing} sources of truth.

This means it is now difficult for ordinary people, and even specialists, to decide what is ``true.''
In current industry parlance, the problem of determining whether a statement is ``true'' is called {\em fact-checking}.
{\em Fact-checking at scale} is the problem of checking {\em all} facts (or, propositions) of interest.
We might say that fact-checking at scale is to fact-checking what ``big data analysis'' is to data analysis.
This is the problem we are going to solve in this paper.

\section{Formulating the {\em Fact-Checking} Problem}
There are two primary ways to formulate the {\em fact-checking problem}. These are the {\em boolean} and {\em relational} formulations.

\subsection{Boolean Fact-Checking}
Boolean fact-checking is the assignment of a boolean label {\em true} or {\em false}, to some given {\em proposition}.
For example, relative to the statement {\em the American economy is strong in 2020}, a boolean system would assign this a binary label of either true or false.
A related variant to this problem is to report a single real-valued {\em probability} (between $0$ and $1$) for the given proposition.
PolitiFact, for example, indicates low probability with the figurative term ``pants-on-fire,'' referring to an idomatic expression for telling a falsehood.

\subsection{Relational Fact-Checking}
Relational fact-checking involves the retrieval of related documents, facts, or propositions, for a given proposition $p$, with some notion of agreement polarity.
For example, relative to the statement {\em the American economy is strong in 2020}, a relational system might find related financial data, and indicate whether they {\em support} or {\em contradict} the original assertion.

\subsection{Conclusion}
We believe that both formulations are interesting.
However, we believe that {\em relational} fact-checking is primary.
As we will see, it is possible to solve relational fact-checking, without first solving boolean fact-checking.
We believe boolean fact-checking can be created on the basis of a relational fact-checking system, using a form of {\em belief propagation}.

\section{Deconstructing the Article-Form Fact-Check}
\label{s:article}
In many popular formulations, including PolitFact, the ``checking of a fact'' is done through the creation of an {\em argument} presented in {\em article-form}.
A {\em article} is a sequence of sentences, perhaps along with images.
An {\em article-form argument} is a sequence of sentences meant to {\em derive} or {\em support} the target proposition $p$ being checked.

\subsection{The Recursive Nature of Fact-Checking}
An {\em article-form fact-check} for the purported ``fact'' (i.e. proposition) $p$ takes the form of an {\em article}, which for us is a sequence of sentences.
This article, in order to make its case will use some of its sentences to introduce {\em assumptions}, and use the rest of the sentences to justify {\em inferences}, based on the assumptions introduced.

In symbolic terms, we can characterize the situation as follows.
We want to argue for the proposition $p$, and we must introduce some assumptions in order to do this, which we will denote $q_1, ..., q_N$.
Thus, in order to check one initial proposition $p$, we end up introducing $N$ additional assumptions.

\subsection{Recursive Explosion}
Because we are interested in fact-checking in the first place, we presumably believe, {\em any fact we are depending on should be checked}.
In such a case, an article-form fact-check raises more questions than it answers.
That is, where we once had one proposition $p$ to check, we now have $N$ new propositions to check after the first round.
But, these $N$ article-form fact-checks will each introduce $N$ new assumptions each, meaning we will have $N^2$ to check after the second round, $N^3$ after the third round, and so on.
Thus, not only does an article-fact check induce more new facts to check than we started with, but it actually induces {\em exponentially} more.
Assuming a relatively constant staff, an employee-driven fact-checking organization will generate facts to check exponentially faster than they can check them.

\subsection{The Base Case Problem}
The second problem we have with recursion is that there is no inherent ``base case'' to a recursive fact-checking operation.
The checking of one fact requires the checking of $N$ others.
Does the recursive growth in the number of relevant facts grow forever?
What happens when a fact-check for a proposition $p$ ends up depending on itself?
These philosophical questions admit, we believe, no known answer other than the {\em scientific method} itself.
This will require us to recast the problem from one of checking ``facts'' to one of checking {\em theories}.

\section{The Scientific Method}
The scientific method presents a coherent and contradiction-free approach to the pursuit of determining ``truth'' and evaluating ``facts.''
It does this by checking entire {\em theories}, rather than individual facts.
A theory is a set of propositions.
In order to leverage the coherent and  well-developed philosophical basis of the scientific method, we will base our platform on the concepts behind the scientific method.

\subsection{Logical Calculus and Logical Theories}
Underlying the application of science we must first have a {\em language} in which to write our propositions, and a set of {\em inference rules} to draw conclusions based on initial assumptions.
The pair of language and inference rules are called a {\em logical calculus}.
We can use for a logical calculus any descendant of the {\em propositional calculus} \citep{andrews:86}.
For example, we can use the propositional calculus itself, or the {\em first-order calculus}, or an {\em intensional calculus}.
The calculus specifies a {\em language} $L$ and a set of {\em inference rules} $R$.
$L$ can be seen as a denumerably infinite set of propositions $L = \left\{p_1, p_2, ... \right\}$.
A {\em theory} $T$ (in $L$) is a {\em set} of propositions, each taken from $L$.

\subsection{Data and Explanations}
In science, propositions can be categorized as either {\em data} or {\em explanatory principles}.
{\em Data} record measurements of the ``outside world.''
Aside from data, we have {\em explanatory principles}, which attempt to ``explain'' the data, by {\em predicting} it, as a consequence of a theory.
We generally prefer the theory ``does the most with the least,'' in the sense that it predicts as much data as possible, while also being as small as possible (cf. Occam's Razor).
Now, the data are just measurements, and as such can be {\em incorrect}.
Sometimes, scientists will try to argue against the accuracy of data, rather than give up on a theory.
Nevertheless, incorrect data is still data, and must be distinguished from explanatory principles.

\subsection{Inference}
Underlying the scientific method is a {\em calculus}, which we underline, is a descendant of the {\em propositional calculus}.
Aside from introducing a {\em language}, the calculus a set of {\em inference rules}.
The inference rules allow us to extend a theory $T$.
An inference rule $R(t_1, ..., t_m)\rightarrow \alpha$, says that, if $\left\{t_1, ..., t_m\right\} \subseteq T$, then we can create a new set $T' = T \cup \left\{\alpha\right\}$, where $\alpha$ is {\em inferred} from $T$. 
For example, from assuming {\em all men are mortal} and {\em Socrates is a man}, we can infer {\em Socrates is mortal}, in first-order logic.

\subsection{Contradiction}
Because we have assumed we are working with a {\em propositional calculus} we have the concept of {\em negation}.
The negation of the statement $\alpha$ would be the statement {\em $\alpha$ is not true}, written $\neg \alpha$.
If a theory contains $\alpha$ and $\neg \alpha$, it is said to be {\em inconsistent}.
A theory that is inconsistent is also said to have a {\em contradiction}.
If there are no contradictions, the theory is called {\em consistent}.
From this perspective, the goal of science is to produce theories consistent with the data, and not inconsistent with it.
One can argue against a theory either by contradicting directly its core assumptions, or else by contradicting one of its predictions. 

\subsection{The Universe of all Theories}
An individual theory must be consistent to be correct.
However, the set of all possible theories, if unified together, would not be consistent.
That is, at any given time, there groups of active theories that cannot all be true.
We can regard the set of all theories, or else all actively considered theories, as a set of theories called the {\em universe of active theories}.
The union of all propositions in all active theories can be called the {\em universe of all active propositions}.
Our goal then is to be able to ``check'' any proposition in the universe of active propositions, and do so without degrading in the (normal) case that the union of all active theories cannot consistently be merged.

\section{The Power of Platforms}
\subsection{The Notion of a ``Platform''}
There has been much discussion recently in various kinds of press, including business and political, about the notion of a ``platform.''
We define a {\em platform} as a {\em channel} that connects {\em many producers} to {\em many consumers}.
That is, each producer can user the channel to some set of consumers.
The producer and consumer can be in {\em different} organizations, in which case the constitute {\em provider} and {\em client}.
Or the producer and consumer can be in the {\em same} organizations, in which case they constitute two steps in the {\em supply chain}.

Under this definition, an old-fashioned physical marketplace would be kind of a platform, because it connects many producers to many consumers at the same time.
A highway network is a platform, because it transports goods from producer to consumer, as is a railroad network.

In the information-technology field, one platform is built upon another.
The personal-computer hardware platform connects software producers to consumers.
The operating system sits atop the hardware, and is a platform that connects application producers to consumers.
The Internet browser sits atop the operating system, and is a platform connecting web site producers to consumers.
Popular web sites like Amazon, Google and Facebook sit atop the Internet browser and the Internet, and connect users to other users.

\subsection{Information-Sharing Platforms}
An {\em information-sharing platform} is a platform whose object is the communication, storage and organization of {\em information}.\footnote{
By ``information'' we can either mean arbitrary sequences of bits.
Claude Shannon famously studied the quantification of {\em the amount} of information in a sequence of bits, but we do not need to worry here about how to quantify information.
}
Information platforms are interesting because they allow us as humans to transcend the normal bounds that exist on how many individuals can work effectively together.

Dunbar's number suggests that the largest number of meaningful relationships a person can have is $150$ \citep{dunbar:92}.
The reason hypothesized for a limit is that each person has a limited information-processing capacity.
In order to interact with someone meaningfully, we not only need to spend time communicating with them, but also considering what they have said once communication is over.
Using information-sharing platforms, like Google Search, Facebook, Twitter, etc., we are able to effectively work with, communicate with, and otherwise share information with far more people than had every been possible before.
Information-sharing platforms allow us to create super organizations, that can achieve more than organizations without the same technology. 

The ability for people to share information more effectively than ever before has precipitated the crisis of fact-checking, requiring the need for fact-checking at scale.
However, we can also turn this crisis to opportunity, by using the power of the platform itself to solve the problem of fact-checking.

\section{The ``PolitiFact'' Model}
PolitiFact is an {\em employee-based organization} that produces {\em article-form fact-checks}.
There are a variety of different groups that do such article-form ``fact-checking'', such as {\em PolitiFact}, {\em Snopes}, {\em CNN}, and {\em The Washington Post}.
For illustrative effect, we use PolitiFact as an exemplar to stand for the whole group.
We will show how the PolitiFact model is inferior to a {\em platform model} of fact-checking.

\subsection{Definitions}
Let us break down the component concepts involved in our claim.

\subsubsection{Employee-Based Organization}
An {\em employee-based organization}, is a fixed legal organization, whose work is done by legal employees.
Each employee has a fixed work day.
There are two central problems that plague a fixed organization.

{\em Limited Staff}:
First, it is difficult to hire and maintain a highly qualified staff. These difficulties grow at least quadratically as the organization grows.
We believe empirically staff can scale by at most $O(\sqrt{N})$ compared to the number of users $N$.

{\em Limited Decision-Making}:
Decision-making is even more limited in an organization than is its entire staff.
For the most important decisions, top-level management will have to be involved.
At the least, usually senior mangement will have to be involved.
We might say that decision-making capacity is ultimately $O(1)$, in the sense that there  is constant bound on the number of individuals that can be considered ``senior management.''

\subsubsection{Article-Form}
The second part of the notion of {\em employee-based article-form fact-checking organization} is that the organization produces {\em article-form} fact-checks.
We reviewed in \S \ref{s:article}, why an article-form fact-check introduces more questions than it answers.
Every proposition which is checked, introduces $N$ new assumptions that need checking.
{\em This creates an exponential growth in the number of facts of interest}, even starting with a single initial fact of interest.
The problem is worse by a constant factor with more initial facts of interest.

\subsection{Problems}
Here, then, are our criticisms of the PolitiFact model.

\subsubsection{Scale}
We have seen in \S \ref{s:article}, that each article-form fact-check introduces more questions than it raises, leading to {\em exponential growth} in the number of facts of interest, having begun with one fact of interest.
Relative to the growth in the number of facts of interest, the organization can only grow at $O(1)$, because no new employees are automatically hired and trained simply by the realization of a need to check more facts.
And, the decision-making bandwidth of an organization is limited by senior management, which can ever only grow to size $O(1)$.
Thus, {\em the research bandwidth and decision-making bandwidth of a fact-checking organization can never scale with the  number of facts of interest}.

\subsubsection{Narrow Bias and Lack of Inclusion}
Historically, we have observed that individual fact-checking organizations are accused of having an overly {\em narrow bias}.
That is, the opinions of the organization do not admit of certain groups' views.
In other words, the narrowness of the bias, makes users feel like their opinions are not {\em included}.
We see here that the perception of a narrow bias is a necessary consequence of the use of an employee-baed fact-checking organization.
This is because the organization can take $O(1)$ opinions per $2^N$ many possible configurations.
Obviously $O(\frac{1}{2^N})$ an exponentially small fraction.
Thus, the fraction of views expressed is a {\em narrow} fraction, and necessarily so.

\section{A Platform-Based Formulation of Fact-Checking}
The employee-based model of fact-checking falls short, as we saw, on the problems of scale, bias, and inclusion.
In contrast, a platform-based solution would have none of these short-comings, as it easily incorporates many people, of diverse biases, and allows literally all users to be included.
The only question then is, {\em how} can we formulate fact-checking to work with platforms.

\subsection{Arguments For-and-Against}
The answer is that we will operationalize the task of {\em fact-checking}, as the task of giving the {\em best arguments for-and-gainst} a given proposition $p$.
This is, in other words, a primarily {\em relative} fact-checking formulation.

\subsection{Messages and their Metadata}

\subsubsection{Broadcast Messages}
The user-facing interface will allow each user to post a {\em broadcast message}, in which that user communicates a public message to potentially all other users.
Theoretically, we think of messages as falling into two groups: {\em propositional nodes} and {\em proof nodes}.
Propositional nodes introduce individual propositions (assumptions), and proof nodes derive new propositions from previously proven ones, by use of argumentation.  

\subsubsection{Comments and Agreement Polarity}
A comment $c^m$ for message $m$ is also a message, but with certain metadata.
The notation $c^m$ is used to write that message $c$ is a {\em reference to} message $m$.
The notation $c^m_a$ is used to write that message $c$ has {\em agreement polarity} $a$ with respect to message $m$.
There are three possible {\em agreement polarities}: 1) {\em agree}, 2) {\em disagree}, and 3) {\em no opinion}.
These are represented internally in our software by the values $1$, $0$ and $null$ respectively.

\subsubsection{Hotness Score}
Each user $u$ can indicate a {\em hotness score} for a given message $m$, indicated by $\sigma(u, m)$.
Hotness is a synonym for {\em high energy}.
To say that a post is hot is to say, relative to whether you agree with it, whether or not it is an {\em important} post, and whether it should rank highly in people's searches.
This score is constrained to be within a finite range, so let us assume for simplicity $0 \leq \sigma(u, m) \leq 1$ for all $u$, $m$.
The users input these scores through the user interface.

\subsubsection{Two-Dimensional Ratings}
Twitter, Reddit, YouTube and Instragram allow the user to rate a post in effectively one dimension.
Either the posts is {\em liked}, or {\em disliked}.
We have now introduced two different dimensions that, we claim, are not being distinguished on these existing platforms.
In the first dimemsion, the user indicates or not they {\em agree} with a message.
In the second dimension, the user indicates whether or not it is {\em hot}.
This corresponds to {\em direction} and {\em magnitude} of a {\em velocity} vector.
{\em To our knowledge, this is the first work to propose to use two different information dimensions in user ratings.}\footnote{
Facebook allows the user to select from a variety of emoticons, but we believe these emoticons do not produce an actionable multi-dimensional semantic meaning.
If they do, they do not capture our agreement-hotness dichotomy.
}

\subsection{Relative Fact-Checking Query}
The crucial new query, {\em that has not been possible} with one-dimensional quality ratings is then: {\em list me the best arguments for (or against) the proposition $p$}.

{\bf Algorithm}:
To run this query, we first select all comments $c^p_a$ that {\em refer to} $p$ with {\em agreement polarity} $a$.
That is, we {\em filter} these comments according to their {\em agreement polarity}.
Finally, we sort the remaining comments by {\em hotness}.
This way, we have got the best arguments, either for or against, the target proposition $p$.

This formulation allows us to present a version of {\em relative fact-checking}, which is the task of, for a proposition $p$, bringing up relevant other information, and indicating its relation to the proposition $p$.

\subsection{Theorem-Proving Completeness}
In this section we will show how to construct a logical calculus {\em within our network formulation}, which is equivalent in power to the theorem-proving system of the propositional calculus.
That is, whatever can be proven in propositional calculus can be proven in our platform calculus.
In fact, we can show that each node will have $O(1)$ size in the length of the proof overall.
This means that ours is not a trivial formulation, in which, e.g., an entire proof is simply written in a single message.
And, by construction, whatever {\em cannot} be proven in the propositional calculus cannot be proven in our calculus either.
The propositional calculus is {\em consistent} and {\em complete}, meaning that a set of hypotheses $H$ derives a proposition $\alpha$, if and only if the inference is really warranted.
{\em Thus, our new calculus is also consistent and complete} for the propositional calculus.

\subsubsection{Proof}
We base our {\em platform calculus} on the simple proof system from \citep{andrews:86}.
We choose this formulation of propositional logic for simplicity, because it relies on a single inference rule, called {\em modus ponens}.
Modus ponens is the rule that, based on assumptions $\alpha$ and $\alpha \rightarrow \beta$, allows us to conclude $\beta$.

Suppose we have a set of assumptions $H$, and we want to use this to prove some proposition.
Andrews' system allows us to create a proof in which each line is either:
1) an assumption, $h \in H$, from the hypotheis of interest,
2) an axiom from a special set of logical truisms, $\gamma \in A$, or 
3) a statement derived from two earlier statements by an application of {\em modus ponens}.
This is analagous to creating a graph fragment, that grows by one node on each iteration.

We can create a proof system in our platform as follows.
We divide messages into two types: {\em propositional nodes} and {\em proof nodes}.
First, we identify each node $h$ in the hypothesis $H$ with a propositional node.
Then, to build proofs on top of these hypothesis nodes, we introduce the {\em proof nodes}.
In logical terms, an application of modus ponens must have two inputs (i.e., some $\alpha$ and $\alpha\rightarrow\beta$).
In terms of our node graph, a proof node can either refer to two hypothesis nodes, one hypothesis node and one logical truism, or else two logical truisms.
The truism assumptions are written right on the node that uses them.
There are at most two truisms per node, thus, there is a $O(1)$ propositions that need to be written on a proof node, relative to the size of the proof.

Using this construction, we can translate any proof in Andrews' consistent-and-complete system, into our construction.
So, anything that can be proven in Andrews' system can also be proven in our system.
QED.

\subsection{Personalization}
The ordinary {\em DimensionRank} algorithm describes how to give a personalized search result based on a non-personalized result \citep{coppola:20}.
That is, to personalize a search result for user $u$, we simply take the non-personalized search result, $R$, and tailor this set to $u$ using their unique user embeddings, to obtain $R^u$, personalized to $u$.
We can of course apply the same principle here to the special case to give a {\em personalized ranking of arguments for-and-against}.
In other words, one user might be interested in certain arguments more than another, and each user would be able to get a personalized experience using this algorithm.

\subsection{Authoritative Voices}
Bill Gates has recently suggested that one of the most important questions is how to balance {\em authoritative voices} with the activities of the {\em citizen-users} of the platform \citep{gates:20}.

{\em We can offer a solution that balances the public's desire to express themselves, with the additional desire to maintain an ability to allow ``authoritative voices,'' wherever they may come from, to weigh in.}
Suppose we have authoritative voices, whose voices we want to promote for a given fact. 
We can simply promote their comments to any target message $m$ to occur as a {\em prefix} to the arguments from the rest of the platform.
To maintain a balance towards free speech and civic involvement, the rest of the users still have the ability to offer alternative arguments, both for and against, any target message.
Finally, both platform users and designated authoritative voices can each recursively fact-check one another.
This is a possibility that hasn't been possible before our platformization of the fact-check.

\subsection{Implementation}
\label{s:impl}
We have implemented this algorithm in our open-source DimensionRank search and social media software package, available at \mbox{\em thinkdifferentagain.art}.
We have our own network running this algorithm at \mbox{\em deeprevelations.com}.

\section{Conclusion}
We have presented what we believe is the first workable proposal for fact-checking at scale, through the platformization of fact-checking.
We gave a detailed analysis of the short-comings of an {\em employee-based fact-checking organization}, which we called the ``PolitiFact model.''
We have released this software and have our own implementation as described in \S \ref{s:impl}.

\bibliography{emnlp-ijcnlp-2019}

\begin{thebibliography}{4}
\expandafter\ifx\csname natexlab\endcsname\relax\def\natexlab#1{#1}\fi

\bibitem[{Andrews(1986)}]{andrews:86}
Peter~B. Andrews. 1986.
\newblock \emph{An Introduction to Mathematical Logic and Type Theory: To Truth
  Through Proof}.
\newblock Academic Press, Inc., Orlando, Florida.

\bibitem[{Coppola(2020)}]{coppola:20}
Gregory Coppola. 2020.
\newblock \href {http://arxiv.org/abs/arXiv:2005.13007} {Dimensionrank:
  Personal neural representations for personalized general search}.

\bibitem[{Dunbar(1982)}]{dunbar:92}
R.~I.~M. Dunbar. 1982.
\newblock Neocortex size as a constraint on group size in primates.
\newblock \emph{Journal of Human Evolution}, 22(6):469--493.

\bibitem[{Gates(2020)}]{gates:20}
Bill Gates. 2020.
\newblock \href
  {https://www.wsj.com/video/watch-bill-gates-at-wsj-ceo-council-summit/031C310C-128B-4229-BC1D-660B99BDA759.html}
  {Bill {G}ates at {WSJ} {CEO} council summit}.
\newblock \emph{Wall Street Journal}.

\end{thebibliography}
\bibliographystyle{acl_natbib}

\appendix

\end{document}